\documentclass[reqno,a4paper,11pt]{article}
\usepackage{color}

\usepackage{epsfig,psfrag,graphicx}
\usepackage[textwidth = 430 pt, textheight = 630 pt]{geometry}

\definecolor{MyDarkBlue}{rgb}{0.15,0.25,0.45}
\usepackage{epstopdf}
\usepackage{amsmath,amssymb}
\usepackage{amsfonts}
\usepackage{mathrsfs}
\usepackage{bbm}

\usepackage{latexsym}
\usepackage{amsthm}

\usepackage[utf8x]{inputenc}

\usepackage{hyperref}
\hypersetup{
hypertexnames=false,
colorlinks=true,
citecolor=MyDarkBlue,
linkcolor=MyDarkBlue,
urlcolor=MyDarkBlue,
pdfauthor={Simone Rea and Christian S\"amann},
pdftitle={The Phase Diagram of Scalar Field Theory on the Fuzzy Disc},
pdfsubject={hep-th math-ph}
breaklinks=true
}

\let\fn\footnote
\renewcommand{\footnote}[1]{\linespread{1.1}\fn{#1}\linespread{1.29}}

\makeatletter\renewcommand{\section}{\@startsection
{section}{1}{\z@}{-3.5ex plus -1ex minus
    -.2ex}{2.3ex plus .2ex}{\bf }}
\makeatletter\renewcommand{\subsection}{\@startsection{subsection}{2}{\z@}{-3.25ex
plus -1ex minus
   -.2ex}{1.5ex plus .2ex}{\it }}
\makeatletter\renewcommand{\subsubsection}{\@startsection{subsubsection}{3}{-2.45ex}{-3.25ex
plus -1ex minus -.2ex}{1.5ex plus .2ex}{\it }}
\renewcommand{\thesection}{\arabic{section}}
\renewcommand{\thesubsection}{\arabic{section}.\arabic{subsection}}
\renewcommand{\@seccntformat}[1]{\@nameuse{the#1}.~~}

\renewcommand{\theequation}{\thesection.\arabic{equation}}
\makeatletter \@addtoreset{equation}{section}

\usepackage[toc,page]{appendix}

\newtheorem{thm}{Theorem}[section]
\renewcommand{\thethm}{\thesection.\arabic{thm}}

\newtheorem{remark}[thm]{Remark}

\renewcommand{\appendices}{
\section*{Appendix}\label{appendices}\setcounter{subsection}{0}
\addcontentsline{toc}{section}{Appendix}
\setcounter{equation}{0}
\makeatletter
\renewcommand{\theequation}{\Alph{subsection}.\arabic{equation}}
\renewcommand{\thesubsection}{\Alph{subsection}}
\renewcommand{\thethm}{\Alph{subsection}.\arabic{thm}}
\@addtoreset{equation}{subsection}
\@addtoreset{thm}{subsection}
\makeatother
}

\usepackage[all,knot]{xy}
\xyoption{arc}

\def\slasha#1{\setbox0=\hbox{$#1$}#1\hskip-\wd0\hbox to\wd0{\hss\sl/\/\hss}}

\def\periodb#1{\setbox0=\hbox{$#1$}#1\hskip-\wd0\hbox to\wd0{-}}

\newcommand{\unit}{\mathbbm{1}}   			

\newcommand{\CCI}{\mathscr{I}}

\newcommand{\CH}{\mathcal{H}}

\newcommand{\CI}{\mathcal{I}}

\newcommand{\CO}{\mathcal{O}}

\newcommand{\CZ}{\mathcal{Z}}

\newcommand{\FR}{\mathbbm{R}}     			
\newcommand{\FC}{\mathbbm{C}}     			
\newcommand{\NN}{\mathbbm{N}}     			
\newcommand{\CPP}{{\mathbbm{C}P}}    			

\newcommand{\dd}{\mathrm{d}}     			
\newcommand{\dpar}{\partial}     			
\newcommand{\de}{\mathrm{e}}     			
\newcommand{\di}{\mathrm{i}}     			
\newcommand{\eps}{{\varepsilon}}			

\newcommand{\eand}{{\qquad\mbox{and}\qquad}}     		

\newcommand{\der}[1]{\frac{\dpar}{\dpar #1}}   		
\newcommand{\tr}{\,\mathrm{tr}\,}     			

\newcommand{\sU}{\mathsf{U}}     			

\newcommand{\sEnd}{\mathsf{End}}

\renewcommand{\remark}[1]{}     				
\def\tyng(#1){\hbox{\tiny$\yng(#1)$}}			
\def\tyoung(#1){\hbox{\tiny$\young(#1)$}}			

\newcommand{\beq}{\begin{eqnarray}}
\newcommand{\eeq}{\end{eqnarray}}

\begin{document}
\begin{titlepage}
\begin{flushright}
 EMPG--15--11
\end{flushright}
\vskip 2.0cm
\begin{center}
{\LARGE \bf The Phase Diagram of \\[0.4cm] Scalar Field Theory on the Fuzzy Disc}
\vskip 1.5cm
{\Large Simone Rea and Christian S\"amann}
\setcounter{footnote}{0}
\renewcommand{\thefootnote}{\arabic{thefootnote}}
\vskip 1cm
{\em Maxwell Institute for Mathematical Sciences\\
Department of Mathematics, Heriot-Watt University\\
Colin Maclaurin Building, Riccarton, Edinburgh EH14 4AS, U.K.}\\[0.5cm]
{Email: {\ttfamily sr227@hw.ac.uk , C.Saemann@hw.ac.uk}}
\end{center}
\vskip 1.0cm
\begin{center}
{\bf Abstract}
\end{center}
\begin{quote}
Using a recently developed bootstrapping method, we compute the phase diagram of scalar field theory on the fuzzy disc with quartic even potential. We find three distinct phases with second and third order phase transitions between them. In particular, we find that the second order phase transition happens approximately at a fixed ratio of the two coupling constants defining the potential. We compute this ratio analytically in the limit of large coupling constants. Our results qualitatively agree with previously obtained numerical results. 
\end{quote}
\end{titlepage}

\section{Introduction}

By a fuzzy space, one usually means a geometric quantization of a compact K\"ahler manifold. The compactness of the space implies that the arising Hilbert space and therefore also the algebra of observables given by the endomorphisms of this Hilbert space are finite dimensional. Correspondingly, there is a minimal resolution with which the space is perceived, which renders it ``fuzzy.'' For this reason, fuzzy spaces are good candidates for regularizing quantum field theories \cite{Grosse:1995ar}, because the path integral over all observables turns into a finite number of ordinary integrals. To compare to what extent quantum field theories on fuzzy spaces provide an approximation to the corresponding quantum field theories, it is particularly useful to study the phase diagrams of the theories in the thermodynamic limit. 

To this end, we need to evaluate the free energy of the fuzzy quantum field theory. As usual in geometric quantization, real functions on compact K\"ahler manifolds are mapped to hermitian operators, representing the quantization on the corresponding fuzzy manifold. Therefore, scalar field theories on fuzzy spaces are simply hermitian matrix models. Contrary to the hermitian matrix models most common in the literature, however, these matrix models come with a kinetic term containing fixed external matrices. This kinetic term presents an obstacle for applying the usual techniques for solving matrix models directly. In particular, the action is no longer invariant under similarity transformations and diagonalizing the matrix is no longer straightforward. This problem can be overcome by rewriting the kinetic term as a multitrace expression, which can then be solved at least in the limit of large matrix size, e.g.\ by the saddle point approximation. 

This approach was developed in \cite{O'Connor:2007ea} and used there, in \cite{Saemann:2010bw} and in \cite{Ihl:2010bu} to compute the phase diagram of scalar quantum field theory on fuzzy complex projective spaces as well as on a three-dimensional space consisting of the Cartesian  product of $\FR$ and the fuzzy sphere. Further applications of this technique are found in \cite{Ydri:2014uaa}, see also \cite{Nair:2011ux,Polychronakos:2013nca,Tekel:2014bta}. The rewriting of the kinetic term was done by applying techniques from group theory, making the calculations somewhat cumbersome. An alternative bootstrapping method for turning the kinetic term into multitraces was then found in \cite{Saemann:2014pca}. Here, enough conditions on the coefficients in the multitrace expressions are derived to fix them uniquely.

The purpose of this letter is to use the bootstrapping approach to compute the phase diagram of scalar field theory on the fuzzy disc \cite{Lizzi:2003ru} and to compare the result to the numerical findings of \cite{Lizzi:2012xy}. The fuzzy disc is particularly appealing as the kinetic term is somewhat simpler than in the case of the fuzzy sphere. One may therefore hope that on the fuzzy disc, quantum scalar field theory is better behaved than on the fuzzy sphere or even that the resulting hermitian matrix model is fully solvable. 

This letter is structured as follows. In section 2, we briefly review the construction of the fuzzy disc and the definition of quantum scalar field theory on it, fixing our conventions. Section 3 deals with rewriting the resulting action as a multitrace expression and taking the limit of large matrix size. We compute the phase diagram of the model using a saddle point approximation in section 4, where we also compare our results to the numerical literature. We conclude in section 5.

\section{The model}

We start our discussion with a concise review of scalar field theory on the fuzzy disc.

\subsection{Fuzzy disc}

The fuzzy disc \cite{Lizzi:2003ru} provides a quantization of the algebra of functions on the unit disc in $\FC\cong\FR^2$. It is obtained by truncating and rescaling the matrix algebra of the Moyal plane. Moreover, one can obtain the fuzzy sphere by gluing together two fuzzy discs. Below, we briefly recall its construction, following roughly \cite{Lizzi:2003ru}.

Recall that the Moyal plane $\FR^2_\theta$ is the geometrical quantization of the complex plane with respect to the canonical symplectic structure, see e.g.\ \cite{IuliuLazaroiu:2008pk} for details. The result of this procedure is an infinite dimensional Hilbert space $\CH$ which agrees with the usual Hilbert space of the harmonic oscillator up to a normalization. That is, we have a vacuum state $|0\rangle$ together with annihilation and creation operators satisfying
\begin{equation}
 \hat{a}|0\rangle=0~,~~~ [\hat a, \hat a^\dagger]=\theta~,~\theta\in \FR^{> 0}~.
\end{equation}
We denote the eigenstates of the number operator $\hat N=\hat a\hat a^\dagger$ with eigenvalue $n\theta$ by $|n\rangle$:
\begin{equation}
 \hat N |n\rangle=n\theta|n\rangle~,~~~n\in \NN~.
\end{equation}
We also introduce a corresponding basis for $\CH^*$: $\langle m|$, $m\in \NN$, normalized such that $\langle m|n\rangle=\delta_{mn}$. The endomorphisms on $\CH$ are spanned by linear combinations of the operators $|m\rangle \langle n|$. The Berezin symbol map $\sigma$ arising in the quantization assigns to each element of $\sEnd(\CH)$ a function on $\FC\cong \FR^2$, providing a dequantization map. We have
\begin{equation}
 \hat{f}=\sum_{m,n=0}^\infty f_{mn}|m\rangle \langle n|~,~~~\sigma(\hat f)(z,\bar z)=\de^{-\frac{|z|^2}{\theta}}\sum_{m,n=0}^\infty f_{mn} \frac{\bar z^m z^n}{\sqrt{m!n!\theta^{m+n}}}~,
\end{equation}
where $z\in \FC$.\footnote{We follow physics conventions and write $f(z,\bar z)$ for a non-holomorphic function $f$.} In particular, as the usual quantization axioms demand, $\sigma(\unit)=1$. Note that real functions correspond to the hermitian endomorphisms $\sEnd_H(\CH)$.

Let us now introduce the following projector on a sub-Hilbert space $\CH_\circ$:
\begin{equation}
 P_N:=\sum_{n=0}^{N-1}|n\rangle \langle n|~,~~~\CH_\circ:= P_N\,\CH\eand \sEnd(\CH_\circ):=P_N\,\sEnd(\CH)\,P_N~.
\end{equation}
The Berezin symbol of this projector reads as
\begin{equation}
 \sigma(P_N)(z,\bar z)=\sum_{n=0}^{N-1}\frac{r^{2n}}{n!\theta^n} \de^{-\frac{r^2}{\theta}}=\frac{\Gamma(1+N,\frac{r^2}{\theta})}{\Gamma(1+N)}~,~~~r=\sqrt{\bar z z}~.
\end{equation}
Here, $\Gamma(n,x)$ is the incomplete gamma function. Note that
\begin{equation}
 \lim_{N\rightarrow \infty}(\sigma(P_N)(z,\bar z))=\left\{\begin{array}{ll}
                                                           1 & \mbox{for}~~\frac{r^2}{\theta}<N~,\\
                                                           0 & \mbox{for}~~\frac{r^2}{\theta}>N~,
                                                          \end{array}\right.
\end{equation}
which shows that the projector $P_N$ corresponds to a step function with support on the disc $D_R\subset \FC$ of radius $R=\sqrt{N \theta}$. This justifies the identification of the algebra $\sEnd_H(\CH_\circ)$ with a quantization of the algebra of real functions on $D_R$, and we call the corresponding noncommutative space the {\em fuzzy disc}. We will always work with a disc of radius $R=1$, fixing $\theta=\frac{1}{N}$.

\subsection{Scalar field theory on the fuzzy disc}

To study scalar field theory on the fuzzy disc, we have to introduce a Laplace operator, i.e.\ an operator on $\sEnd_H(\CH_\circ)$, which approximates the usual Laplace operator $\Delta=4\der{z}\der{\bar z}$. Recall that geometric quantization and the Berezin symbol map lead to the identification
\begin{equation}
 N[\hat a,-]\sim \der{\bar z}\eand -N[\hat a^\dagger,-]\sim \der{z}
\end{equation}
on the Moyal plane. On the fuzzy disc, we can combine these operators with the projectors $P_N$ to obtain a Laplace operator. There are two obvious candidates:
\begin{equation}
 \Delta_N \hat f:=-4N^2P_N[\hat a,[\hat a^\dagger,\hat f]]P_N\eand \Delta_N \hat f:=-4N^2P_N[\hat a,P_N[\hat a^\dagger,\hat f]P_N]P_N
\end{equation}
for $\hat f\in \sEnd_H(\CH_\circ)$. The first one was used in \cite{Lizzi:2003ru} and \cite{Lizzi:2012xy}, but the second one has the advantage that $\Delta_N \unit=0$, as one might expect for the constant function $\sigma(\unit)(z,\bar z)=1$. The latter expectation is somewhat debatable as constant functions on the disc are in fact step functions, and one might argue that the fuzzy boundary should lead to deviations from $\Delta_N \unit=0$. In the following, we will nevertheless work with the second Laplace operator for two reasons. First, this choice simplifies our computations dramatically and second, we will be mostly interested in the large $N$ limit, in which both choices agree anyway.

The second ingredient we need is the notion of an integral. Geometric quantization and normalization of the integrals lead to the following identification:
\begin{equation}
 \int \frac{\dd^2 z}{2}~ \sigma(\hat f)(z,\bar z)=\frac{\pi R^2}{N}\tr(\hat f)=\pi\theta\tr(\hat f)=\frac{\pi}{N}\tr(\hat f)~,
\end{equation}
where the last equality is again due to our choice $R=\sqrt{N\theta}=1$.

Introducing the shorthand $\check a:=P_N \hat a P_N$, we can now write down the action for scalar field theory on the fuzzy disc with quartic potential terms:
\begin{equation}\label{eq:action}
S[\hat \Phi] =  \frac{\pi}{N} \tr\left(-4N^2\hat \Phi [ \check a, [ \check a^{\dagger}, \hat \Phi ] ] +r\hat \Phi^2 + g\hat \Phi^4 \right)~,
\end{equation}
where $r,g\in \FR$ such that $rx^2+gx^4$ is bounded from below for all $x\in\FR$. From now on, we will regard $\hat{\Phi}\in\sEnd_H(\CH_\circ)$ as a hermitian $N\times N$-matrix $\Phi$ and drop the hat to simplify our notation. The partition function for the action \eqref{eq:action} is then given by
\begin{equation}\label{eq:partition_function}
 \CZ:=\int \dd \mu_D(\Phi)~\de^{-\beta S[\Phi]}~,
\end{equation}
where $\dd \mu_D(\Phi)$ denotes the usual Dyson measure on the space of hermitian matrices.

\section{Computing the partition function}

As in the cases of the fuzzy sphere and fuzzy $\CPP^n$ in general \cite{O'Connor:2007ea,Saemann:2010bw,Saemann:2014pca}, one cannot apply the usual techniques of hermitian matrix models to the partition function \eqref{eq:partition_function} in a straightforward manner. This is due to the fact that the kinetic term presents an obstacle to a simple diagonalization of $\Phi$. To circumvent this problem, we will rewrite the kinetic term as multitrace expressions.

\subsection{Multitrace action}

As usual in dealing with hermitian matrix models, we wish to diagonalize the hermitian matrix $\Phi$ as $\Phi=\Omega\Lambda\Omega^\dagger$, where $\Lambda$ is a diagonal matrix containing the eigenvalues $\lambda_1,\ldots,\lambda_N$ of $\Phi$ and $\Omega\in \sU(N)$. Under this change of variables, the Dyson measure $\dd \mu_D(\Phi)$ on the space of hermitian $N\times N$-matrices, which appeared in the partition function \eqref{eq:partition_function}, decomposes according to
\begin{equation}
\int \dd \mu_D(\Phi) \ =\ \int \prod_{i=1}^N \dd \lambda_i~ \Delta^2(\Lambda) \int \dd \mu_H(\Omega)~.
\end{equation}
Here, $\Delta(\Lambda)$ is the Vandermonde determinant:
\begin{equation}
\Delta(\Lambda) \ := \ \det([\lambda_i^{j-1}]_{ij}) \ =\ \prod_{i>j}{(\lambda_i-\lambda_j)}
\end{equation}
and $\dd \mu_H(\Omega)$ is the Haar measure on $\sU(N)$. In a partition function, the Vandermonde determinant induces a repulsive interaction between eigenvalues:
\begin{equation}\label{eq:partion_VDM}
\begin{aligned}
\CZ =  \ &\int \prod_{i=1}^N \dd \lambda_i ~ \Delta^2(\Lambda) \int \dd \mu_H(\Omega)~ \de^{-\beta S[\Omega\Lambda\Omega^\dagger]}\\
=  \ &\int \prod_{i=1}^N \dd \lambda_i ~ \int \dd \mu_H(\Omega)~\de^{-\beta S[\Omega\Lambda\Omega^\dagger] +2\sum_{i>j}{\log|\lambda_i-\lambda_j|}}~.
\end{aligned}
\end{equation}

In the case of ordinary hermitian matrix models, the action is invariant under $\Phi\rightarrow \Omega\Phi\Omega^\dagger$. Therefore, the integral over the Haar measure just gives a constant factor. In the case of our model \eqref{eq:action}, however, the kinetic term $S_{\rm kin}[\Phi] = -4\pi N \tr \! \left(\Phi[\check a ,[\check a^{\dagger},\Phi]]\right)$ is not invariant under this transformation and hence our first goal is to compute the following integral:
\begin{equation}
 \CCI:=\int \dd \mu_H(\Omega)~\de^{\,\beta\,4\pi N \tr \! \left(\Omega\Lambda\Omega^\dagger[\check a ,[\check a^{\dagger},\Omega\Lambda\Omega^\dagger]]\right)}
\end{equation}
As shown in \cite{O'Connor:2007ea}, it is possible to rewrite the kinetic term as a sum of traces and multitraces of polynomials in $\Phi$ under the integral. This multitrace action is then invariant under $\Phi\rightarrow \Omega\Phi\Omega^\dagger$ and the integral over $\Omega$ becomes trivial.

Since our model is invariant under $\Phi \to - \Phi$ , our multitrace action will only contain terms of even total power of $\Phi$. At each order $\alpha$, there are $p(\alpha)$ terms in $S_{\rm MT}$ of total power $\alpha$ in $\Phi$, where $p(\alpha)$  denotes the number of integer partitions of $\alpha$. We label these coefficients by $a_{\pi_1,\pi_2,\ldots,\pi_k}$, where $\pi_1+\pi_2+\ldots+\pi_k$ is a partition of $\alpha$:
\begin{equation}\label{eq:MT-action}
\begin{aligned}
 S_{\rm MT}[\Phi]=&a_2\tr(\Phi^2)+a_{1,1}\tr(\Phi)^2+a_4\tr(\Phi^4)+a_{3,1}\tr(\Phi^3)\tr(\Phi)+\\
 &+a_{2,2}\tr(\Phi^2)^2+a_{2,1,1}\tr(\Phi^2)\tr(\Phi)^2+a_{1,1,1,1}\tr(\Phi)^4+\ldots~.
\end{aligned}
\end{equation} 
Two methods have been developed to compute the coefficients $a_{\pi_1,\ldots,\pi_k}$. The first one uses group theoretic techniques and was applied in  \cite{O'Connor:2007ea} and \cite{Saemann:2010bw} to compute the partition function of scalar field theory on fuzzy $\CPP^N$. The second one is a bootstrapping method presented in \cite{Saemann:2014pca}, which is more robust and more easily implemented, and we will use this method in the following.

The basic idea behind the bootstrapping method consists in choosing suitable differential operators $D$ such that
\begin{equation}
\begin{aligned}
D \, \de^{-\beta S_{\rm kin}[\Phi]}\ =\ & D \, \de^{-\beta S_{\rm MT}[\Phi]}~,\\
D\,\de^{-\beta S_{\rm kin}[\Phi]} = \CO_{\rm kin}[\Phi]\,\de^{-\beta S_{\rm kin}[\Phi]}~~~\mbox{and}&~~~D\,\de^{-\beta S_{\rm MT}[\Phi]} = \CO_{\rm MT}[\Phi]\,\de^{-\beta S_{\rm MT}[\Phi]}~,
\end{aligned}
\end{equation}
and both $\CO_{\rm kin}$ and $\CO_{\rm MT}$ are invariant under $\Phi \rightarrow \Omega\Phi\Omega^\dagger$ for $\Omega \in \sU(N)$. Then the operators $\CO_{\rm kin}$ and $\CO_{\rm MT}$ can be pulled out of the integral and the equation
\begin{equation}\label{bootstr}
 D \,\int \dd \mu_H(\Omega)~\de^{-\beta S_{\rm kin}[\Phi]} = D \,\int \dd \mu_H(\Omega)~\de^{-\beta S_{\rm MT}[\Phi]}
\end{equation}
implies $\CO_{\rm kin}=\CO_{\rm MT}$. The left hand side of \eqref{bootstr} only depends on $\beta$ and $N$ while the right hand side depends on $\beta$, $N$ and the coefficients $a_{\pi_1,\ldots,\pi_k}$. Hence, given a sufficient number of differential operators, equations  \eqref{bootstr} will yield enough conditions to fix all the coefficients in $S_{\rm MT}[\Phi]$, thereby giving the desired rewriting of the action.

It was found in \cite{Saemann:2014pca} that the operator $\sum_a \der{\Phi_{aa}} =: \der{\Phi_{aa}}$ yields conditions that fix more than half the unknown coefficients in $S_{\rm MT}[\Phi]$. More precisely, we have:
\begin{equation}
 \der{\Phi_{aa}} \, \de^{-\beta S_{\rm kin}[\Phi]} = 16 \beta \pi N \tr\left( [ \check a, [ \check a^{\dagger}, \Phi ] ]\right)~ \de^{-\beta S_{\rm kin}[\Phi]} = 0~,
\end{equation}
from which it follows that
\begin{equation}
\der{\Phi_{aa}}  \, \de^{-\beta S_{\rm MT}[\Phi]} =0~.
\end{equation}
This equation holds, in fact, for any scalar field theory on any fuzzy space if the kinetic term of the continuum theory vanishes on constant functions and the quantization condition $\sigma(\unit)=1$ is fulfilled. This is the case for our choice of Laplace operator on the fuzzy disc.

Equation \eqref{bootstr} for $D=\der{\Phi_{aa}}$ yields $p(\alpha-1)$ conditions on the coefficients of $ S_{\rm MT}[\Phi]$, which we can use to express all coefficients of the form $a_{\pi_1,\ldots,\pi_{k-1},1}$ in terms of other coefficients. In particular, consider the terms in $\CO_{\rm kin}[\Phi]$ and $\CO_{\rm MT}[\Phi]$ corresponding to the partition $\pi_1+\ldots+\pi_{k-1}=\alpha-1$. In terms of coefficients appearing in $S_{\rm MT}$, $\CO_{\rm kin}[\Phi]=\CO_{\rm MT}[\Phi]$ amounts to \cite{Saemann:2014pca}
\begin{equation}
\begin{aligned}
 a_{\pi_1,\pi_2,\ldots,\pi_{k-1},1}=-\frac{1}{rN}\sum_{\sigma}\Big(&(\sigma(\pi_1)+1)a_{\sigma(\pi_1)+1,\sigma(\pi_2),\ldots,\sigma(\pi_{k-1})}+\\
 &+(\sigma(\pi_2)+1)a_{\sigma(\pi_1),\sigma(\pi_2)+1,\ldots,\sigma(\pi_{k-1})}+\ldots\\
 &\hspace{1cm}+(\sigma(\pi_{k-1})+1)a_{\sigma(\pi_1),\sigma(\pi_2),\ldots,\sigma(\pi_{k-1})+1}\Big)~,
\end{aligned}
\end{equation}
where the sum runs over all permutations of $\pi_1,\ldots,\pi_{k-1}$ and $r-1$ is the number of parts $\pi_i$ which are 1. Moreover, we define $a_{\pi_1,\pi_2,\ldots,\pi_{k-1}}:=0$ unless $\pi_1\geq\pi_2\geq\ldots\geq\pi_{k-1}$. In particular, we have
\begin{equation}
a_{1,1} = -\frac{a_2}{N}
\end{equation}
at second order and
\begin{equation}
a_{3,1} = -\frac{4a_4}{N} \ ,\qquad a_{2,1,1} =  \frac{6a_4}{N^2}-\frac{2a_{2,2}}{N} \ , \qquad a_{1,1,1,1} = -\frac{3a_4}{N^3} + \frac{a_{2,2}}{N^2}
\end{equation}
at fourth order. Hence, $S_{\rm MT}[\Phi]$ is determined up to fourth order in $\Phi$ by $a_2$, $a_4$ and $a_{2,2}$. To fix these, we need to turn to higher order differential operators. 

Unfortunately, none of the higher order differential operators yield functionals $\CO_{\rm kin}[\Phi]$ which are $\sU(N)$-invariant in general. However, simply evaluating the result at $\Phi = 0$ gives the desired invariance. To fix $a_2$, we use $D := \der{\Phi_{ab}}\der{\Phi_{ba}}$ and we readily compute
\begin{equation}
\begin{aligned}
 \left. \der{\Phi_{ab}}\der{\Phi_{ba}} \, \de^{-\beta S_{\rm kin}[\Phi]} \right|_{\Phi=0} &=8\pi\beta N^2(N-1)~,\\
 \left. \der{\Phi_{ab}}\der{\Phi_{ba}} \, \de^{-\beta S_{\rm MT}[\Phi]} \right|_{\Phi=0} &= -2\beta (N^2-1) a_2~,
\end{aligned}
\end{equation}
which implies
\begin{equation}
a_2 = -\frac{4\pi N^2}{N+1}~.
\end{equation}

Determining $a_4$ and $a_{2,2}$ is slightly more involved. We define two differential operators $D_1$ and $D_2$,
\begin{equation}
D_1 := \der{\Phi_{ab}}\der{\Phi_{bc}} \der{\Phi_{cd}}\der{\Phi_{da}}\eand
D_2 := \der{\Phi_{ab}}\der{\Phi_{ba}} \der{\Phi_{cd}}\der{\Phi_{dc}}~,
\end{equation}
and we solve the pair of simultaneous equations
\begin{equation} \left.D_i \, \de^{-\beta S_{\rm kin}[\Phi]}\right|_{\Phi=0} \ =\ \left.D_i \, \de^{-\beta S_{\rm MT}[\Phi]}\right|_{\Phi=0}  \ , \qquad i=1,2~,
\end{equation}
which yields
\begin{equation}
 \begin{aligned}
a_4 &= \frac{8  \beta \pi^2 N (12 - 3 N^2 - N^3) }{  3 (N+1)(N+2)(N+3)}~,\\
a_{2,2} &= \frac{ 8\beta \pi^2 (36 + 36 N - 3 N^2 - 11 N^3 - 2 N^4) }{  3 (N+1)^2(N+2)(N+3)}~.
 \end{aligned}
\end{equation}

The computations for the sixth order coefficients are straightforward but cumbersome and we simply quote the result of some computer algebra:
\begin{equation}
 \begin{aligned}
  a_6&=\scalebox{0.93}{$\frac{-64\beta^2\pi^3(N^6-15 N^5-5 N^4-123 N^3+388 N^2-30 N-120)}{3(N-5) (N-3) (N+1) (N+2) (N+3) (N+4) (N+5)}$}~,\\
  a_{4,2}&=\scalebox{0.93}{$\frac{-64\beta^2\pi^3(2 N^{10}+4 N^{9}-103 N^{8}-81 N^7+1462 N^6+1610 N^5-8783 N^4-5865 N^3+10830 N^2+2700 N+3600)}{3(N-5) (N-3) (N-2)N (N+1)^2 (N+2) (N+3) (N+4) (N+5)}$}~,\\
  a_{3,3}&=\scalebox{0.93}{$\frac{64 \beta^2\pi^3(2 N^{10}+2 N^{9}-98 N^{8}-129 N^7+1634 N^6+1384 N^5-9226 N^4-5905 N^3+13960 N^2+1800 N+2400)}{3 (N-5) (N-3) (N-2)N (N+1)^2 (N+2) (N+3) (N+4) (N+5)}$}~,\\
  a_{2,2,2}&=\scalebox{0.93}{$\frac{64 \beta^2\pi^3(2 N^{9}+6 N^{8}-106 N^{7}-191 N^6+1653 N^5+3008 N^4-9364 N^3-15795 N^2+12135 N+19020)}{3 (N-5) (N-3) (N-2) (N+1)^3 (N+2) (N+3) (N+4) (N+5)}$}~.
 \end{aligned}
\end{equation}
To keep our computations manageable, however, we limit ourselves to multitrace terms of quartic order in $\Phi$. It will turn out that this approximation is sufficient for all our purposes.

\subsection{Limit of large matrix size}

For computing the partition function of the multitrace action \eqref{eq:MT-action} plus the potential term, we turn to the large $N$ limit in order to apply the saddle point approximation later. Note that as usual in quantum field theory, a rescaling of the degrees of freedom needs to be accompanied by a rescaling of the fields and the involved coupling constants. This leads to a multiscaling limit which we discuss now.

As $N$ goes to infinity, the discrete set of eigenvalues $\lambda_1,\ldots, \lambda_N$ of $\Phi$ goes over into a continuous function $\lambda(x)=\lambda(\frac{i}{N})$, $0< x\leq 1$. Sums over powers of eigenvalues, $\tr(\Phi^j)=\sum_i(\lambda_i)^j$ turn into integrals $N\int_0^1\dd x \, \lambda(x)^j$. Note that each trace yields a factor of $N$ when being recast as an integral.

The maximum total scaling of the terms in the action is $\sim N^2$, which is fixed by the exponentiated Vandermonde determinant, cf.\ \eqref{eq:partion_VDM}. The coefficients of the multitrace action scale as follows:
\begin{equation}
a_2 \sim N~,~~~a_4\sim \beta N~,~~~a_{2,2}\sim \beta~.
\end{equation}
We denote the scaling of the eigenvalues $\lambda$, and the couplings $\beta$, $r$, $g$ by $\rho_\lambda$, $\rho_\beta$, $\rho_r$ and $\rho_g$, respectively:
\begin{equation}
\beta =N^{\rho_\beta} \tilde{\beta}~, \qquad \lambda = N^{\rho_\lambda} \tilde{\lambda}~, \qquad r = N^{\rho_r} \tilde{r}~, \qquad g = N^{\rho_g}  \tilde{g} ~.
\end{equation}
With this notation, we obtain the following inequalities:
\begin{equation}
\begin{aligned}
\rho_\beta+2+2\rho_\lambda&\leq 2~,~~~&2\rho_\beta+2+4\rho_\lambda&\leq 2~,~~~&2\rho_\beta+2+4\rho_\lambda&\leq 2~,\\
\rho_\beta+\rho_r+2\rho_\lambda&\leq 2~,~~~&\rho_\beta+\rho_g+4\rho_\lambda&\leq 2~.
\end{aligned}
\end{equation}
Each inequality corresponds to a summand in the action. For each summand, we have one power of $\beta$ outside of the action and further powers from the coefficients, powers of $N$ from the various traces and the coefficients as well as scalings of further couplings and the eigenvalues. We would like to choose a scaling that saturates as many inequalities as possible. The first three inequalities are saturated by $\rho_\beta=-2\rho_\lambda$, and the last two equations yield $\rho_r=2$ and $\rho_g=-2\rho_\lambda$. A convenient choice is therefore
\begin{equation}
 \rho_\lambda=\rho_\beta=\rho_g=0\eand \rho_r=2~.
\end{equation}

We can now write our model in terms of rescaled quantities. Instead of integrating over $x$, we integrate over $\lambda$:
\begin{equation}
 \int_0^1\dd x \rightarrow \int_\CI \dd \lambda\, \rho(\lambda)~,
\end{equation}
where $\rho(\lambda):=\frac{\dd x}{\dd \lambda}$ is the eigenvalue density and $\CI$ is its support. Introducing the moments
\begin{equation}
 c_i:=\int_{\CI}\,\dd \lambda\,\rho(\lambda)\,\lambda^i~,
\end{equation}
we arrive at the action
\begin{equation}\label{eq:large-N-action}
\begin{aligned}
 {\beta} {S}={\beta}\Big(&\tilde{r} c_2+ g  c_4-4\pi( c_2- c^2_1)-\tfrac{8}{3} \beta \pi^2\big( c_4-4 c_3 c_1+6 c_2 c^2_1-3 c^4_1+2( c_2- c^2_1)^2\big)\Big)+\\
 &-\int_\CI\,\dd \lambda\,\dd \mu\,\rho(\lambda)\log(|\lambda-\mu|)\rho(\mu)~.
\end{aligned}
\end{equation}

\section{The phase diagram}

We now come to the computation of the phase diagram of scalar field theory on the fuzzy unit disc in the large $N$ limit. We also compare our result to the numerical findings of \cite{Lizzi:2012xy}. 

\subsection{Saddle point approximation}

To compute the action \eqref{eq:large-N-action}, we rewrite it as 
\begin{equation}\label{eq:large-N-action-V}
S[\rho(\lambda)] = \int_\CI\,\dd \lambda\,\rho(\lambda) V(\lambda)-\int_\CI\,\dd \lambda\,\dd \mu\,\rho(\lambda)\log(|\lambda-\mu|)\rho(\mu)+\xi \left( \int_\CI\,\dd \lambda\,\rho(\lambda)-1\right)~,
\end{equation}
where the Lagrange multiplier $\xi$ was included to fix the normalization of the eigenvalue density. The potential reads as 
\begin{equation}
 V(\lambda)=\beta  \left(\alpha_4 \lambda ^4+\alpha_{31} c_1 \lambda ^3+\lambda ^2 \left(\alpha_2+\alpha
_{211} c_1^2+\alpha_{22} c_2\right)+\lambda  \left(\alpha_{1111} c_1^3+\alpha_{11}
   c_1\right)\right)~.
\end{equation}
This potential is in fact the general potential for the rewritten action of a quantum scalar field theory on an arbitrary fuzzy space\footnote{assuming that the Laplace operator satisfies $\Delta \unit=0$}, truncated at fourth order in $\Phi$. We recover our action \eqref{eq:large-N-action} with the following choice of coefficients:
\begin{equation}
\begin{aligned}
\alpha_{11}&=4 \pi~,~~~&\alpha_{1111}&=\frac{8 \pi ^2 \beta }{3}~~~,~&\alpha_2&=\tilde{r}-4 \pi~,~~~&\alpha_{211}&=-\frac{16 \pi ^2 \beta }{3}~,\\
\alpha_{22}&=-\frac{16 \pi ^2 \beta }{3}~,~~~&\alpha_{31}&=\frac{32 \pi ^2 \beta }{3}~,~~~&\alpha_4&=g-\frac{8 \pi ^2 \beta }{3}~.
\end{aligned}
\end{equation}

The saddle point equation is obtained by varying the action \eqref{eq:large-N-action-V} with respect to $\rho(\lambda)$:
\begin{equation}\label{eq:eom}
\tilde V(\lambda)-2\int_\CI\,\dd \mu\,\rho(\mu)\log(|\lambda-\mu|)+\xi = 0~,
\end{equation}
where 
\begin{equation}
\begin{aligned}
 \tilde V(\lambda)=\beta\Big(&\alpha_4\lambda^4+\alpha_{31}(c_1\lambda^3+c_3\lambda)+\alpha_2\lambda^2+\alpha_{211}c_1^2\lambda^2+\\
 &+2\alpha_{211}c_2c_1\lambda+2\alpha_{22}c_2\lambda^2+4\alpha_{1111}c_1^3\lambda+2\alpha_{11}c_1\lambda\Big)~.
\end{aligned}
\end{equation}

The key object in finding the solution to \eqref{eq:eom} is the resolvent
\begin{equation}
 W(\lambda):=\int \dd \mu~\frac{\rho(\mu)}{\lambda-\mu}~,
\end{equation}
which is an analytic function on $\FC\backslash \CI$. A detailed review of the application of the resolvent is given e.g.\ in \cite{Brezin:1977sv,DiFrancesco:1993nw}, but for our purposes, the following observations are sufficient. First, note that we expect essentially two cases: $\CI$ can either be given by a single interval or by the disjoint union of two intervals. We will refer to these cases as the single cut and the double cut solutions. The resolvent's singular part $\omega(\lambda)$ contains two roots $\delta_i$ in the former case and four roots in the latter case. It necessarily satisfies
\begin{equation}\label{eq:polynomial_matching}
 \omega^2(\lambda)=M^2(\lambda)\prod_i(\lambda-\delta_i)=\tilde V'{}^2(\lambda)-R(\lambda)~,
\end{equation}
where $M(\lambda)$ is some polynomial in $\lambda$ and $R(\lambda)$ is a polynomial of one degree less than $\tilde V'(\lambda)$. The jump over the cut $\CI$ yields the eigenvalue density according to
\begin{equation}
 \rho(\lambda)=-\frac{1}{2\pi\di}\big(W(\lambda+\di \eps)-W(\lambda-\di \eps)\big)~,
\end{equation}
which implies that
\begin{equation}
 \rho(\lambda)=\frac{1}{2\pi}\,|M(\lambda)|\,\sqrt{(\delta_2-\lambda)(\lambda-\delta_1)}~,~~~\CI=[\delta_1,\delta_2]
\end{equation}
for the single cut solution or, for the double cut solution, 
\begin{equation}
 \rho(\lambda)=\frac{1}{2\pi}\,|M(\lambda)|\,\sqrt{(\delta_4-\lambda)(\lambda-\delta_3)(\delta_2-\lambda)(\delta_1-\lambda)}~,~~~\CI=[\delta_1,\delta_2]\cup[\delta_3,\delta_4]~,
\end{equation}
where we assumed $\delta_4>\delta_3>0>\delta_2>\delta_1$.

\subsection{Solutions}

As explained later, we will be interested in two types of solutions. The eigenvalue densities of these solutions have support on a single interval and the disjoint union of two intervals, respectively. The results of related computations can be found in \cite{Shimamune:1981qf,Shishanin:2007zz} as well as \cite{O'Connor:2007ea,Saemann:2010bw}.

We will start with the former case and put $\CI=[s-d,s+d]$. Equation \eqref{eq:polynomial_matching} fixes the polynomial
\begin{equation}
 M(\lambda)=m_0+m_1\lambda+m_2\lambda^2
\end{equation}
as well as $d$. We then use the normalization of $\rho(\lambda)$, $c_0=1$, and the self-consistency conditions for $c_1$, $c_2$ and $c_3$ to fix the remaining unknowns. We obtain the eigenvalue density
\begin{equation}
 \rho(\lambda)=\frac{1}{2\pi}\,|m_0+m_1\lambda+m_2\lambda^2|\,\sqrt{d^2-(s-\lambda )^2}
\end{equation}
with
\begin{equation}
\begin{aligned}
&m_0=\beta(2 \alpha_{211} c_1^2+3 s
   \alpha_{31} c_1+2 \alpha_2+2(d^2+2 s^2)
   \alpha_4+4 c_2 \alpha_{22})~,\\
 &\hspace{2.2cm}m_1=\beta(4 s\alpha_4+3c_1\alpha_{31})~,~~~m_2=4\beta\alpha_4~.
\end{aligned}
\end{equation}
Additionally, we obtain from \eqref{eq:polynomial_matching} the equation
\begin{subequations}\label{eq:constraints_asym_single_cut}
\begin{equation}
 \begin{aligned}
 &8 \alpha_4 s^3+6 c_1 \alpha_{31} s^2+\left(12 \alpha_4 d^2+4 \alpha_2+8 c_2
   \alpha_{22}+4 c_1^2 \alpha_{211}\right) s+4 c_1 \alpha_{11}+\\
   &\hspace{4cm}+3 d^2 c_1 \alpha
_{31}+2 c_3 \alpha_{31}+4 c_1 c_2 \alpha_{211}+8 c_1^3 \alpha_{1111}=0~.
 \end{aligned}
\end{equation}

The normalization condition $c_0=1$ implies
\begin{equation}
\tfrac{1}{4} \beta  d^2 \left(2 \alpha_2+2 \alpha_{211} c_1^2+4 \alpha_{22} c_2+6 \alpha_{31}
   c_1 s+3 \alpha_4 \left(d^2+4 s^2\right)\right)=1~,
\end{equation}
and self-consistency conditions for $c_1$, $c_2$ and $c_3$ read as 
\begin{equation}
 \begin{aligned}
8 \beta  d^2 s \left(\alpha_2+2 \alpha_{22} c_2+3 \alpha_4 \left(d^2+2 s^2\right)\right)+\hspace{4cm}&\\
+c_1 \left(\beta  d^2 \left(8 \alpha_{211} c_1 s+3 \alpha_{31} \left(d^2+8
   s^2\right)\right)-16\right)&=0~,\\
 \end{aligned}
\end{equation}
\begin{equation}
 \begin{aligned}
    2 c_2 \left(\alpha_{22} \beta  d^2 \left(d^2+4 s^2\right)-4\right)+\beta  d^2 \Big(\alpha
_{211} c_1^2 \left(d^2+4 s^2\right)+\hspace{3cm}\\
   +6 \alpha_{31} c_1 s \left(d^2+2 s^2\right)+\alpha_2\left(d^2+4 s^2\right)+2 \alpha_4 \left(d^4+12 d^2 s^2+12 s^4\right)\Big)&=0~,\\
 \end{aligned}
\end{equation}
and
\begin{equation}
 \begin{aligned}
32 c_3+\beta  d^2 \Big(-4 \alpha_{211} c_1^2 \left(3 d^2 s+4 s^3\right)-4 \alpha_2 \left(3 d^2
   s+4 s^3\right)-8 \alpha_{22} c_2 s \left(3
   d^2+4 s^2\right)+\\
   -3 \alpha_{31} c_1 \left(d^4+18 d^2 s^2+16 s^4\right)-12 \alpha_4 s \left(3 d^4+14 d^2 s^2+8 s^4\right)\Big)&=0~.
 \end{aligned}
\end{equation}
\end{subequations}
The combined solution of these equations leads to rather involved expressions which do not provide any further insight. The special case $s=0$, however, does yield manageable expressions and it will be of interest to us later on. Here, the eigenvalue distribution is symmetric and correspondingly, $c_1=c_3=0$. This also implies that $\alpha_{211}$, $\alpha_{31}$ and $\alpha_{1111}$ can be put to zero. We obtain the eigenvalue density
\begin{equation}
 \rho(\lambda)=\frac{\sqrt{d^2-\lambda ^2}~\left|4-d^2 \beta  \left(d^2-4 \lambda ^2\right)
   \alpha_4\right|}{2 d^2 \pi }~,
\end{equation}
where the boundary $d$ is determined by the equation
\begin{equation}
 \beta ^2 \alpha_4 \alpha_{22} d^8+\left(12 \beta  \alpha_4+4 \beta  \alpha_{22}\right) d^4+8 \beta  \alpha_2 d^2-16=0~.
\end{equation}
Note that the above equations can be used to reproduce the results of \cite{Saemann:2010bw} with the appropriate choices of $\alpha_2$, $\alpha_{22}$ and $\alpha_4$.

Let us now turn to the symmetric double cut solution with support on the interval $\CI=[-\sqrt{s+d},-\sqrt{s-d}]\cup[\sqrt{s-d},\sqrt{s+d}]$. Again, this is a solution with a symmetric eigenvalue density, and we thus put $c_1=c_3=\alpha_{211}=\alpha_{31}=\alpha_{1111}=0$. Following the same steps as above, we obtain 
\begin{equation}
\rho(\lambda) =\frac{2 \alpha_4 \beta  |\lambda|  \sqrt{d^2-\left(s-\lambda ^2\right)^2}}{\pi }~.
\end{equation}
The normalization condition of the eigenvalue density and the self consistency for $c_2$ lead to 
\begin{equation}
d^2\beta \alpha_4 =1~,~~~d^2\beta \alpha_4 s = c_2~,
\end{equation}
which, together with \eqref{eq:polynomial_matching}, yield
\begin{equation}
c_2 = s~,~~~d=\frac{1}{\sqrt{\beta \alpha_4}}~,~~~s= -\frac{\alpha_2}{2(\alpha_4+\alpha_{22})}~.
\end{equation}

\subsection{Phase diagram}

Comparing with previous studies of scalar field theories on fuzzy spaces \cite{O'Connor:2007ea,Saemann:2010bw} as well as the numerical results of \cite{GarciaFlores:2005xc,Panero:2006bx,GarciaFlores:2009hf,Lizzi:2012xy}, we expect three phases: a symmetric single cut solution, in which the eigenvalue density has support on a single symmetric interval $\CI=[-d,d]$, a symmetric double cut solution, in which the eigenvalue density has support on two symmetric intervals $\CI=[-\sqrt{s+d},-\sqrt{s-d}]\cup[\sqrt{s-d},\sqrt{s+d}]$ and an asymmetric single cut solution, in which the eigenvalue density has support on the interval $\CI=[s-d,s+d]$. These phases are also called {\em disorder phase}, {\em non-uniform order phase} and {\em uniform order phase}, respectively. There should be a third order phase transition between the first two phases and a second order transition between the last two phases. The former is actually the usual phase transition in a hermitian matrix model with quartic even potential, while the latter is the analogue of the usual phase transition in two-dimensional scalar quantum field theory, cf.\ \cite{Glimm:1975tw}.

We start by considering the existence boundaries of the various phases. We are interested in the parameter range in which a phase transition can occur, i.e.\ essentially positive $g$ and negative $r$. Note, however, that for the potential to be bounded from below, we need a positive coefficient of $\lambda^4$ in the potential \eqref{eq:large-N-action-V}. This restricts our parameter space to
\begin{equation}
 \tilde r\leq 0 \eand g>\frac{8\pi^2\beta}{3}~.
\end{equation}

At the existence boundary of the symmetric single cut solution, the eigenvalue density develops a third root at $\lambda=0$, signaling the transition to the symmetric double cut regime. Putting $\rho(0)=0$, we can solve for $d^2$, which, together with the consistency condition $c_0=1$ yields an expression for $c_2$, which in turn yields the existence boundary
\begin{equation}\label{eq:eb_sym_single_cut}
 \alpha_2=-\frac{2(\alpha_4+\alpha_{22})}{\sqrt{\beta \alpha_4}}~.
\end{equation}
In the case of our model \eqref{eq:large-N-action}, the existence boundary is thus at
\begin{equation}
 \tilde r=\frac{2 \left(8 \pi ^2 \beta-g \right)}{\sqrt{3\beta 
   \left(3 g-8 \pi ^2 \beta \right)}}+4 \pi~.
\end{equation}

Next, we turn to the existence boundary of the symmetric double cut solution supported on the interval $\CI=[-\sqrt{s+d},-\sqrt{s-d}]\cup[\sqrt{s-d},\sqrt{s+d}]$. Here, one readily finds that the existence boundary agrees with that of the symmetric single cut, \eqref{eq:eb_sym_single_cut}.

Finally, the asymmetric single cut solution only makes sense as long as $s-d>0$ and $\rho(\lambda)\geq 0$. This leads to complicated algebraic relations, which cannot be brought into a nice analytical expression. Instead, we simply check the validity of our solutions manually, whenever required.

As shown using an approximation in \cite{Saemann:2010bw}, the asymmetric single cut solution as well as the symmetric double cut solution exist on overlapping regions of the parameter space. To determine the preferred phase, we have to compare the free energy of both solutions since a physical system will adopt the phase with the lowest possible free energy. The latter is defined as $F:=-\log(\CZ)$, where $\CZ$ is the partition function of our model. In the saddle point approximation, we correspondingly have $F=\beta S[\rho(\lambda)]-\beta S_{\rm free}[\rho_{\rm free}(\lambda)]$, where $S_{\rm free}$ is the free action, truncated at quadratic order in $\Phi$ and $\rho_{\rm free}(\lambda)$ is the corresponding eigenvalue density. By subtracting the free part, we only let the connected diagrams contribute to the free energy. Note that 
\begin{equation}
 \beta S[\rho(\lambda)]=\int_\CI \dd \lambda\,\rho(\lambda)\big(V(\lambda)-\tfrac12 \tilde V(\lambda)\big)-\tfrac12 \xi~,
\end{equation}
as follows from \eqref{eq:large-N-action-V} and \eqref{eq:eom}. The Lagrange multiplier $\xi$ is determined by solving \eqref{eq:eom} at a suitable point $\lambda\in \CI$. For example, in the case of the symmetric single cut solution, we can choose $\lambda=0$ to obtain $\xi=2\int_\CI\dd \mu\,\rho(\mu)\log|\mu|$ as well as the following expression for the free energy:
\begin{equation}
 F=\tfrac{3}{128} \beta ^2 \alpha_4^2 d^8+\tfrac{1}{32} \beta ^2 \alpha_2 \alpha_4
   d^6-\tfrac{1}{8} \beta  \alpha_4 d^4+\tfrac{1}{8} \beta  \alpha_2
   d^2-\tfrac{1}{2}\log \left(-\tfrac{1}{2} d^2 \beta  \alpha
_2\right)-\tfrac14
\end{equation}
or
\begin{equation}
 \begin{aligned}
  F=&\tfrac{1}{384} \beta ^2 \left(3 g-8 \pi ^2 \beta \right)^2 d^8+\tfrac{1}{96} (\tilde{r}-4
   \pi ) \beta ^2 \left(3 g-8 \pi ^2 \beta \right) d^6-\tfrac{1}{24} \beta  \left(3
   g-8 \pi ^2 \beta \right) d^4+\\
   &+\tfrac{1}{8} (\tilde{r}-4 \pi ) \beta  d^2-\tfrac{1}{2}\log \left(-\tfrac{1}{2} d^2 (\tilde{r}-4 \pi ) \beta \right)-\tfrac14
 \end{aligned}
\end{equation}
in the case of our model \eqref{eq:large-N-action}.

In the cases of the symmetric double cut and asymmetric single cut solutions, we follow the choices of \cite{Saemann:2010bw} and determine $\xi$ at $\lambda=\pm\sqrt{s}$ and $\lambda=s$, respectively. For the symmetric double cut, we have
\begin{equation}
\begin{aligned}
 \beta S[\rho(\lambda)]=\int_\CI\dd{\lambda}\,\rho(\lambda)&\left(V(\lambda)-\tfrac{1}{2}\tilde V(\lambda)-\tfrac12\log|\lambda-\sqrt{s}|-\tfrac12\log|\lambda+\sqrt{s}|\right)+\\
 &\hspace{1cm}+\tfrac{1}{4}\tilde V(\sqrt{s})+\tfrac{1}{4}\tilde V(-\sqrt{s})~,
\end{aligned}
\end{equation}
which evaluates to
\begin{equation}
 F=-\frac{\beta  \alpha_2^2}{4 \left(\alpha_4+\alpha_{22}\right)}+\tfrac{1}{4} \log
   \left(\frac{\alpha_4}{\beta  \alpha_2^2}\right)-\tfrac{3}{8}~,
\end{equation}
and for our model \eqref{eq:large-N-action} reads as
\begin{equation}
 F=-\frac{\beta  (\tilde{r}-4 \pi )^2}{4 \left(g-8 \pi ^2 \beta \right)}+\tfrac{1}{4} \log
   \left(\frac{3 g-8 \pi ^2 \beta }{3 (\tilde{r}-4 \pi )^2 \beta }\right)-\tfrac{3}{8}~.
\end{equation}

For the asymmetric single cut, we have
\begin{equation}
 \beta S[\rho(\lambda)]=\int_\CI\dd{\lambda}\,\rho(\lambda)\left(V(\lambda)-\tfrac{1}{2}\tilde V(\lambda)-\log|\lambda-s|\right)+\tfrac{1}{2}\tilde V(s)~,
\end{equation}
which leads to the following lengthy expression:
\begin{equation}
 \begin{aligned}
  F=&\tfrac{3}{128} \beta ^2 \alpha_4^2 d^8+\tfrac{3}{16} s^2 \beta ^2 \alpha_4^2
   d^6-\frac{3 s^3 \beta ^2 \alpha_4^2 d^6}{8 c_1}+\tfrac{1}{32} \beta ^2 \alpha_2 \alpha_4 d^6+\tfrac{1}{32} \beta ^2 c_1^2 \alpha_4 \alpha_{211}
   d^6+\\
   &-\tfrac{1}{8} \beta  \alpha_4 d^4+\tfrac{1}{8} \beta  \alpha_2
   d^2-\tfrac{3}{2} s^2 \beta  \alpha_4 d^2+s \beta  c_1 \alpha_4 d^2-\frac{s^3
   \beta  \alpha_4 d^2}{c_1}+\tfrac{1}{8} \beta  c_1^2 \alpha_{211} d^2+\\
   &+\frac{3\beta  c_1^2 \alpha_4 \alpha_{211} d^2}{4 \alpha_{22}}-\frac{3 s \beta  c_1
   \alpha_4 \alpha_{211} d^2}{4 \alpha_{22}}+\frac{\beta  c_1^4 \alpha_{211}^2}{2 \alpha_{22}}-\frac{s \beta  c_1^3 \alpha_{211}^2}{2 \alpha_{22}}-\tfrac{1}{2} \log \left(-\tfrac{1}{2} d \beta  \alpha_2\right)+\\
   &-\tfrac{1}{2} s^2 \beta  \alpha_2+s \beta  c_1 \alpha_2-6 s^4 \beta 
   \alpha_4+2 s^3 \beta  c_1 \alpha_4+\frac{4 s^5 \beta  \alpha_4}{c_1}+s \beta
    c_1 \alpha_{11}+s \beta  c_1^3 \alpha_{211}+\\
    &-\tfrac{1}{2} s^2 \beta  c_1^2
   \alpha_{211}+\frac{\beta  c_1^2 \alpha_2 \alpha_{211}}{2 \alpha_{22}}-\frac{s \beta  c_1 \alpha_2 \alpha_{211}}{2 \alpha_{22}}-\frac{3 s^2
   \beta  c_1^2 \alpha_4 \alpha_{211}}{\alpha_{22}}+\frac{3 s^3 \beta  c_1 \alpha_4 \alpha_{211}}{\alpha_{22}}+\\
   &-\beta  c_1^4 \alpha_{1111}+2 s \beta 
   c_1^3 \alpha_{1111}-\tfrac{1}{4}+\frac{s^2}{3 d^2}+\frac{4 s c_1}{3d^2}-\frac{c_1^2 \alpha_{211}}{\alpha_{22} d^2}+\frac{s c_1 \alpha_{211}}{\alpha_{22} d^2}-\frac{2 s^3}{3 c_1 d^2}+\\
   &-\frac{8 s^4}{d^4}+\frac{8 s^3 c_1}{3 d^4}+\frac{8 s c_1^3 \alpha_{211}}{\alpha_{22} d^4}-\frac{16 s^2 c_1^2 \alpha_{211}}{\alpha_{22} d^4}+\frac{8 s^3 c_1 \alpha_{211}}{\alpha_{22}d^4}+\frac{16 s^5}{3 c_1 d^4}~,
 \end{aligned}
\end{equation}
where the variables are subject to the constraints \eqref{eq:constraints_asym_single_cut}.

As consistency checks of our results, we readily verify that the free energies for the symmetric single cut and the symmetric double cut solutions agree at the common existence boundary, where we expect the usual third order phase transition of hermitian matrix models with quartic order potential. Furthermore, the free energy for the asymmetric single cut reduces to that of the symmetric single cut if $s=0$. Finally, the free energies of both the symmetric single cut and the symmetric double cut solutions reduce to those of \cite{Brezin:1977sv,Shimamune:1981qf,Shishanin:2007zz} with the correct choice of parameters.

The phase transition between the asymmetric single cut and the symmetric double cut is now found by equating the corresponding free energies. The resulting equations are again too involved to be reformulated in analytical form, and we find the corresponding solution using simple numerical methods.

\begin{figure}[h]
\hspace{2cm}
\begin{picture}(240,195)
\includegraphics[scale=1]{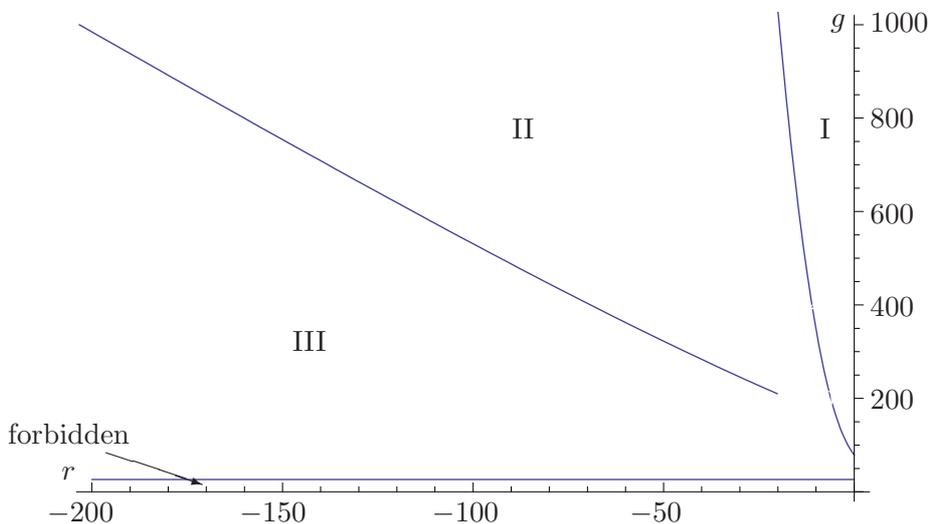}
\put(0,190.0){\makebox(0,0)[l]{$1000$}}
\put(0,154.0){\makebox(0,0)[l]{$800$}}
\put(0,118.0){\makebox(0,0)[l]{$600$}}
\put(0,83.0){\makebox(0,0)[l]{$400$}}
\put(0,48.0){\makebox(0,0)[l]{$200$}}
\put(-308,5){\makebox(0,0)[l]{$-200$}}
\put(-236,5){\makebox(0,0)[l]{$-150$}}
\put(-164,5){\makebox(0,0)[l]{$-100$}}
\put(-90,5){\makebox(0,0)[l]{$-50$}}
\put(-12.0,190.0){\makebox(0,0)[c]{$g$}}
\put(-300.0,20.0){\makebox(0,0)[c]{$r$}}
\put(-17.0,150.0){\makebox(0,0)[c]{I}}
\put(-130.0,150.0){\makebox(0,0)[c]{II}}
\put(-210.0,70.0){\makebox(0,0)[c]{III}}
\put(-286.0,28.0){\vector(3,-1){36}}
\put(-300.0,35.0){\makebox(0,0)[c]{forbidden}}
\end{picture}
\vspace*{-5pt}
\caption{The phase diagram of scalar field theory on the fuzzy disc for $\beta=1$. The phases I, II and III correspond to the symmetric single cut, symmetric double cut and asymmetric single cut solutions. The approximately straight line has asymptotic slope $-\frac{8\pi}{3\sqrt{3}}$.}\label{fig:1}
\end{figure}

Altogether, we obtain the phase diagram given in figure \ref{fig:1}. As discussed above, low values of $g$ are forbidden, which is an artifact of our approximation. Also, for very low values of both $g$ and $|r|$, our approximation of the kinetic term becomes unreliable. In the remaining parameter space, we have three distinct phases in which the symmetric single cut, the symmetric double cut and the asymmetric single cut solutions are appropriate. In figure \ref{fig:1}, these phases are labeled as I, II and III. The phase transition between I and II is the analogue of the usual third order phase transition in hermitian matrix models with quartic even potential. The phase transition between II and III is the second order phase transition also found in real scalar field theory with quartic even potential on the plane. Note that the phase transition from II to III is essentially given by a straight line $g\approx-4.84\,r$, and computing the phase boundary for high values of $-r$ confirm this feature of our plot in figure \ref{fig:1}.

Using our formulas, we can now compute the slope of the straight line analytically. Our approximation of the kinetic term by multitrace expressions of quartic order in $\Phi$ becomes better for large values of $g$ and $r$, and we can restrict ourselves to these. As a second input, we find from numerically solving our equations that at the phase boundary where the free energies of phase II and III agree, $d$ tends to a fixed value $d\approx \tfrac12$. Both assumptions allow us to linearize our equations, which gives us the expressions
\begin{equation}\label{eq:analytical_slope}
 d=\frac{3^{\tfrac14}}{\sqrt{2\pi}}\approx 0.5250\eand g=-\frac{8\pi}{3\sqrt{3}}\,r\approx -4.84\,r~.
\end{equation}

\subsection{Comparison to numerical results}

Having computed the phase diagram, let us now compare our results to those of \cite{Lizzi:2012xy}. There it was found that neither the coupling $g$ nor the field are renormalized, while $r$ is renormalized as $r=\tilde{r}N^{\tfrac13}$. While we also find that only $r$ requires renormalization, our factor is different, $r=\tilde{r}N^2$.

The three phases we obtained are also found in the numerical analysis and as stated before, the symmetric single cut, the symmetric double cut and the asymmetric single cut solutions correspond to the disorder phase, the non-uniform order phase and the uniform order phase, respectively. We recover again the fact that the phase transition between the non-uniform order and the uniform order phases is given by a straight line. However, the best fit of \cite{Lizzi:2012xy} suggested a relation $g\approx -0.51\,r$, while we found $g\approx -4.84\, r$. We are not sure about the source of this discrepancy. It might be due to our different renormalization of $r$ together with errors introduced by our approximation of the kinetic term by expressions of fourth order in $\Phi$ or numerical errors for low values of $N$ in \cite{Lizzi:2012xy}. The fact that we recover a perfect straight line for larger values of $|r|$ suggests that the approximation of the kinetic term is indeed a good one in that parameter range. A last possible cause for the quantitative difference of our results to those of \cite{Lizzi:2012xy} might be the different choices of Laplace operators.

Recall that scalar field theory on $\FR^2$ with quartic even potential exhibits a phase transition with a phase boundary given by a straight line, cf.\ the numerical results of \cite{Loinaz:1997az}. Scalar field theories on two-dimensional fuzzy spaces necessarily inherit this phase transition, and the straight line was found in the numerical studies of \cite{GarciaFlores:2005xc,Panero:2006bx,GarciaFlores:2009hf,Lizzi:2012xy} on the fuzzy sphere and the fuzzy disc. We reproduced this line by analytical methods.

\section{Conclusions}

In this letter, we computed the phase diagram of quantum scalar field theory on the fuzzy disc using bootstrapping and matrix model techniques. These methods are analytical but employ a perturbative expansion of the kinetic term of the scalar field theory, analogous to a high-temperature expansion. The result is a hermitian matrix model with an action containing multitrace terms.

It had been an initial hope that scalar field theory on the fuzzy disc, which is possibly the simplest fuzzy field theory, exhibits some special features allowing a bootstrapping to all orders in the field. Unfortunately, this was not the case.

We proceeded to compute the partition function of the matrix model in the limit of large matrix size using a saddle point approximation. We also derived expressions for the free energy in various phases. All our results were given in a very general form, applying to arbitrary multitrace models with actions of quartic order in the matrix. In particular, the free energy for the various phases of quantum scalar field theories on any fuzzy space can readily be written down using our equations.

We computed the shape of the phase diagram for scalar field theory on the fuzzy disc with quartic even potential and derived the relevant phase boundaries analytically. We find three phases: In the first phase, the disorder phase, the potential has essentially the shape of a single well and the expectation value of the magnitude of the field vanishes. In the second phase, the non-uniform order phase, the potential is a double well and the expectation value of the field distributes symmetrically at the bottom of both wells. In the third phase, the uniform order phase, the potential is a deep double well and the vacuum expectation value of the field equals one of the minima of the well. Our results agree qualitatively with those of \cite{Lizzi:2012xy}, but our quantitative results differ. There may be various reasons for this discrepancy, amongst which are the different renormalization of the couplings as well as errors in the numerical approximations and the different choices of Laplace operators.

The most interesting result of our computations is that we analytically reproduced the linear phase boundary between the non-uniform and the uniform order phases. This is a feature of both ordinary two-dimensional scalar field theory as well as scalar field theory on two-dimensional noncommutative spaces. It was indeed the numerical finding of this linear phase boundary for scalar field theory on the fuzzy sphere in \cite{GarciaFlores:2005xc} which triggered the research of \cite{O'Connor:2007ea}. The original aim there was to find the slope of this phase boundary analytically. In this letter, we finally achieved this goal and for the case of the fuzzy disk, the slope is given in \eqref{eq:analytical_slope}.

\section*{Acknowledgements}
This work was partially supported by the Consolidated Grant ST/L000334/1 from the UK Science and Technology Facilities Council.

\bibliography{bigone}

\bibliographystyle{latexeu}

\end{document}